# Title: Optical biomarker of metabolism for breast tumor diagnosis: Insights from subcellular dynamics

**Authors:** Zichen Yin[1,2]†, Shuwei Zhang[3]†, Bin He[1,2], Houpu Yang[3], Zhengyu Chen[1,2], Zhangwei Hu[1,2], Yejiong Shi[1,2], Ruizhi Xue[1,2], Panqi Yang[1,2], Yuzhe Ying[4], Chengming Wang[1], Shu Wang[3]* & Ping Xue[1,2]*

**Affiliations:**

[1]State Key Laboratory of Low-dimensional Quantum Physics and Department of Physics, Tsinghua University; Beijing, 100084, China

[2]Frontier Science Center for Quantum Information; Beijing, China

[3]Breast Center, Peking University People's Hospital; Beijing 100044, China

[4]Department of Neurosurgery, Beijing Tsinghua Changgung Hospital, School of Clinical Medicine and Institute of Precision Medicine, Tsinghua University; Beijing, 102218, China

† These authors contributed equally to this work

*Corresponding author. Shu Wang email: shuwang@pkuph.edu.cn; Ping Xue email: xuep@tsinghua.edu.cn.

**Abstract:** Label-free metabolic dynamics contrast is highly appealing but difficult to achieve in biomedical imaging. Interference offers a highly sensitive mechanism for capturing the metabolic dynamics of the subcellular scatterers. However, traditional interference detection methods fail to isolate pure metabolic dynamics, as the dynamic signals are coupled with scatterer reflectivity and other uncontrollable imaging factors. Here, we demonstrate active phase modulation-assisted dynamic full-field optical coherence tomography (APMD-FFOCT) that decouples and quantifies the metabolic dynamics by adding a reference movement for all interferential scatterers. This novel technique enables imaging and dynamic analysis of subcellular structures along with their changes during the apoptotic process in tumor tissues. Furthermore, the nucleus-to-cytoplasm dynamic intensity ratio could serve as an optical biomarker for breast tumor grading, enhancing intraoperative diagnosis.

**Main Text:**

**Introduction**

A hallmark of cancer cells is their ability to reprogram their metabolism to support uncontrolled cell growth and proliferation(*1, 2*). Many cancers show deregulated glucose uptake and enhanced glycolytic rates, supporting much higher metabolic rates(*3, 4*). And the malignancy of tumors has proved to be closely related to their metabolism(*5*). Meanwhile, pathological features serve as a gold standard for tumor grading, forming the cornerstone of prognostic predictions and treatment decisions(*6*). However, the standard histopathological process is laborious and time-intensive, involving formalin fixation and paraffin embedding (FFPE), followed by thin sectioning, staining, and mounting on glass slides(*7*). Therefore, imaging techniques capable of delivering rapid, high-resolution histological insights, particularly those that can provide metabolic information, are highly appealing. Such modalities hold the promise of enhancing diagnostic accuracy, enabling real-time monitoring of tissue metabolism, and potentially providing a biomarker of tumor malignancy based on subcellular metabolism.

On a large scale, tumor metabolism imaging, including techniques such as Single Photon Emission Computed Tomography (SPECT)(*8*), Positron Emission Tomography (PET)(*9*), and Magnetic Resonance Spectroscopy (MRS)(*10*), can evaluate the averaged metabolic state of tumors and locate tumors by measuring specific metabolites or radioactive tracers(*11, 12*). However, these techniques are limited in resolution and sensitivity. On a microscopic level, metabolism imaging methods including mass spectrometry-based imaging(*13*), Raman spectroscopy(*14*), and fluorescence microscopy(*15*) enhance the understanding of cellular biochemical processes. These methods enable subcellular spatial mapping of metabolites by detecting their molecular masses, specific molecular vibrations, and multiphoton excitation from both fluorescent dyes and the inherent auto-fluorescence of metabolites(*16, 17*). However, the aforementioned methods mainly obtain metabolic information by measuring the chemical indicator within cells, yet very few approaches are capable of directly imaging the metabolic dynamics. Particle tracking in living cells has provided insight to intracellular dynamics and structures(*18*), however, its low-throughput nature and the necessity for labeling complicate its use, rendering it unsuitable for clinical applications. Currently, there is still a lack of label-free methods for directly imaging the metabolic dynamics from the subcellular to the tissue scale.

Imaging subcellular dynamics within thick tissues presents numerous challenges. Firstly, the spatial scale of intracellular motions typically spans from nanometers to micrometers, and temporally from milliseconds to seconds(*19, 20*), which is beyond the spatial and temporal capture capabilities of most optical imaging instruments(*21*). Secondly, due to the relatively transparent nature of subcellular structures, it is hard to obtain high-contrast images without labeling. Nevertheless, labeling procedures are time-consuming and may potentially interfere with the cells' natural functions[21]. Thirdly, the high density of cells in tissues may lead to overlapping or obscuring of cellular images, making it challenging to distinguish the movement of individual cells. Recently, a novel imaging technique known as dynamic full-field optical coherence tomography (D-FFOCT) has been developed(*22, 23*). This technique employs a Linnik interference microscope to measure temporal fluctuations of back-scattered light, thereby responding sensitively to the nanometer scale dynamics of subcellular scatterers. Therefore, label-free and high-contrast imaging can be achieved by extracting statistical features of interference signals. With enhanced image contrast compared to traditional FFOCT(*24-26*), D-FFOCT has proved highly accurate in diagnosing breast cancer and nodal metastasis(*25, 27*). However, since D-FFOCT was invented, the utilization of metabolic dynamics information is still limited. This is because the fluctuations of interference signals are influenced by multiple coupled factors, such as the dynamics and reflectivity of subcellular scatterers. Moreover, the illumination intensity on the focal plane, the light attenuation in the optical path, and the defocus effect around a depth of interest also significantly affect the detection efficiency and response uniformity of the interference signal(*28*). These factors are sensitive to, for example, the imaging depth and parallelism between the interference plane and the focal plane. Therefore, it is impossible to maintain consistency in all these factors across different images or even different regions of the same image. As a result, the physical interpretation of D-FFOCT images remains ambiguous, and the pure metabolic dynamics cannot be decoupled and quantized. In fact, the inability to decouple pure metabolic dynamics as an independent parameter is a common and critical problem encountered in interference detection, hindering its further utilization and widespread adoption.

In this study, we present a novel technique called active phase modulation-assisted D-FFOCT (APMD-FFOCT), which applies a specially designed oscillation on the reference mirror to actively modulate the interference signal. This active modulation can be treated as a standard motion of all interferential scatterers on the sample arm and therefore can decouple the signal of metabolic dynamics from reflectivity and other uncontrollable imaging factors. We demonstrated that APMD-FFOCT not only can capture subnuclear features like the nucleolus, chromatin, and nuclear envelope of breast tumor cells in three dimensions, but also elucidate their dynamics and reflectivity properties. Moreover, the morphological transformations of cells throughout apoptosis, along with changes in reflectivity and metabolic dynamics, can be continuously imaged and measured over an extended period without the need for labeling. Finally, by analyzing APMD-FFOCT images across various breast tumor grades, we reveal that the dynamic intensity ratio of nucleus to cytoplasm is closely correlated with tumor malignancy and hence propose that this ratio may optically provide a new and fast biomarker of tumor malignancy.

## Results

### APMD-FFOCT system and imaging principle.

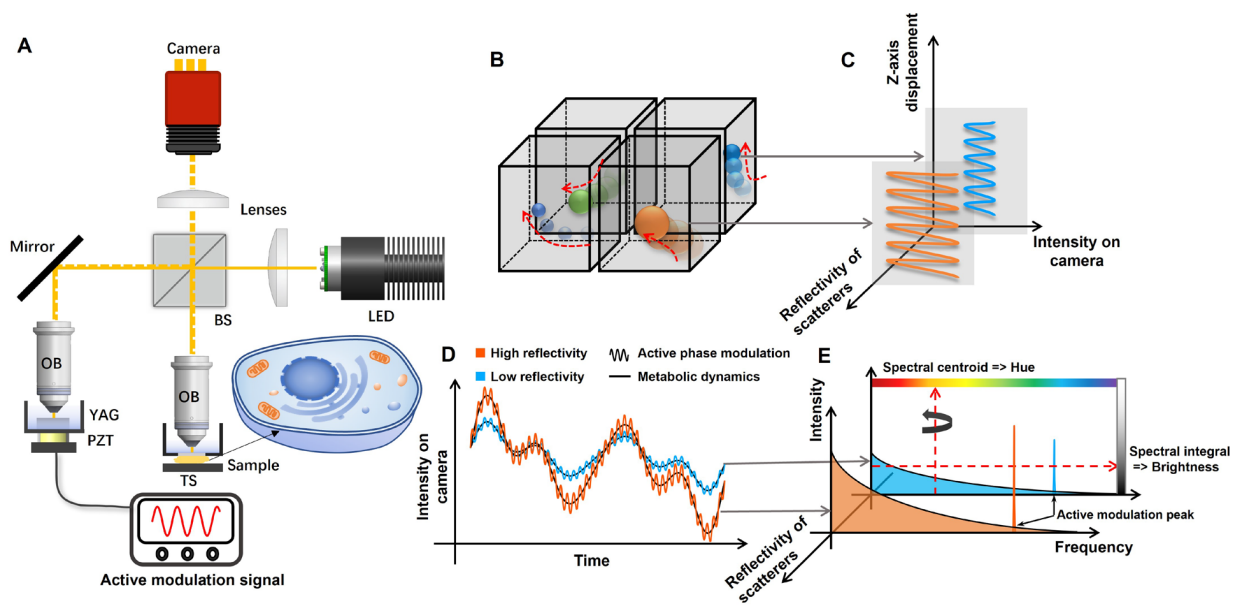

**Fig. 1. APMD-FFOCT system and imaging principle.** (**A**) Schematic of APMD-FFOCT. BS beam splitter; OB water immersion objective; PZT piezoelectric translation; TS translation stage. YAG yttrium

aluminum garnet. The signal generator provides an active modulation signal to drive the PZT in the reference arm. (**B**) Schematic of random walks of subcellular scatterers. The black box corresponds to the detection range of one pixel of the camera. The subcellular scatterers vary in size and composition, causing differences in their reflectivity. (**C**) A schematic illustrating the relationship between z-axis displacement, the reflectivity of scatterers, and light intensity fluctuations on camera, where reflectivity determines the amplitude of light intensity fluctuations. (**D**) Schematic of intensity traces that capture the temporal fluctuations of back-scattered light from two scatterers with differing reflectivity but identical motion. These traces incorporate both the metabolic motions from the sample and the active modulation from the reference mirror. (**E**) Schematic of frequency spectra obtained by Fourier transform of intensity traces. The intensity peaks in the high-frequency regions are caused by the active modulation of the PZT. The low-frequency region in spectra mainly reflects the metabolic dynamics of scatterers. The reflectivity of the scatterers influences both the intensity of the spectral component associated with metabolic dynamics and the height of the active modulation peak.

The APMD-FFOCT setup comprises a Linnik interference microscope with two identical 20x water immersion objectives and a low coherence light source as shown in Fig. 1A. The light returning from the sample layer of interest interferes with the light reflected back from the reference mirror and is subsequently projected onto the camera. The basic principle of APMD-FFOCT is to detect subcellular metabolic movements by capturing the optical length changes of back-scattered light through interferometry(*22*). Additional system details are provided in the Methods section. The tissue on the sample arm is freshly excised, where metabolism is still going on. As shown in Fig. 1B, the subcellular scatterers, which contribute to fluctuations of the interference signals, vary in size and composition, thereby causing differences in their reflectivity. Therefore, the fluctuations of light intensity on the camera are a combined result of z-axis displacement and the reflectivity of scatterers, as illustrated in Fig. 1C. This ambiguous physical meaning of D-FFOCT images hinders its further analysis and application. To solve this problem, we add a sinusoidal signal to the PZT in the reference arm to generate an active phase modulation of the interference signal. The phase modulation of 25 Hz with an amplitude of 23 nm is carefully determined to avoid interfering with the signals of metabolic dynamics that are mainly distributed at the low-frequency region (fig. S1)(*29*). Consequently, the interference signal fluctuations on the camera involve both metabolic motions from the sample and the active modulation from the reference mirror, as shown in Fig. 1D. If we treat the reference mirror as stationary, this active modulation is equivalent to a common oscillation for all interferential scatterers. After the Fourier transform of intensity traces, this active modulation forms sharp intensity peaks on the frequency spectra, as shown in Fig. 1E. Since this active modulation at 25 Hz is consistent for all scatterers and significantly exceeds metabolic dynamics at this frequency, the height of the active modulation peaks in the spectra mainly reflects factors including the reflectivity of scatterers, the illumination intensity and the interference on the focal plane, the light attenuation in the optical path. Therefore, these factors can be eliminated by dividing the frequency spectrum by the height of the active modulation peak. Consequently, pure dynamic intensity of metabolism can be decoupled and extracted as the frequency spectrum integral after the division operation. More theoretical derivation is provided in the Methods section and the experimental verification is shown in fig. S2. It is worth mentioning that the physical meaning of decoupled dynamic intensity can be considered as the contribution of random walk of scatterers in the physical model(*30*), as illustrated in Fig. 1B. In fact, the simulation based on that shows that the dynamic intensity is almost proportional to the velocity of the scatterers, as shown in fig. S3. More details will be discussed in supplementary materials.

The fluctuations of light intensity on each pixel are recorded as a time-domain intensity trace over 5 seconds at a frame rate of 100 FPS, as shown in Fig. 1D. These intensity traces contain a strong static background and a weak fluctuation signal which reflects the motions of the scatterers. Fourier transformation of all the recorded traces then gives the corresponding frequency spectra from the lowest (0.2 Hz) to the highest (50 Hz), which covers most intracellular movements[20, 31]. As shown in Fig. 1E, the hue of the dynamic image pixels is determined by the spectral centroid without the zero-frequency portion, thus reflecting the fluctuation speed[23], transitioning from yellow (indicating slower fluctuations) to blue (indicating faster fluctuations) with green representing intermediate speeds. And the brightness of pixels is determined by the spectral integral excluding the zero-frequency portion, reflecting fluctuation amplitude. Consequently, the dynamic images can be rendered by extracting the dynamic characteristics from frequency spectra.

**The metabolic dynamic intensity of subcellular structures and the dynamic process of apoptosis**

The distribution and morphology of extracellular matrix and tumor cells play a crucial role in tumor diagnosis and grading[32]. Here, we use APMD-FFOCT to image a freshly excised infiltrating ductal carcinoma (IDC, grade III) sample and reveal the dynamic structures of the tumor cells and collagen fibers. Fig. 2A and Fig. 2B are derived from the blue and green portions of the frequency spectrum in Fig. 2D, respectively, representing what can be approximated as reflectivity contrast (verified in fig. S4) and "reflectivity × dynamics" contrast. Since the intensity noise caused by ambient vibration is mainly distributed between 40 Hz and 50 Hz, this region in Fig. 2D denoted by red is eliminated. In the central part of Fig. 2A, there are noticeable dark holes, whose distribution aligns with the cell nuclei as seen in Fig. 2B. This suggests that the reflectivity of these nuclei is lower compared to the surrounding cytoplasm. Intriguingly, despite their lower reflectivity, these nuclei exhibit the strongest intensity in Fig. 2B, indicating very high metabolic dynamics. Moreover, the collagen fibers primarily distribute around the edges of Fig. 2A, which show the highest reflectivity. However, the corresponding region in Fig. 2B shows the lowest intensity, which suggests that tumor-associated collagen fibers, despite being a prevalent highly backscattering structure, exhibit significantly lower metabolic dynamics compared to tumor cells. Considering that the aligned collagen fibers are a prognostic signature for survival in human breast carcinoma[33] but often remain invisible in traditional D-FFOCT, integrating Fig. 2A and Fig. 2B can reveal the distribution and morphology of both collagen fibers and tumor cells, thereby providing a more detailed and comprehensive diagnostic image, as illustrated in fig. S5.

By sequentially capturing dynamic images in the z-direction and filtering out the high-frequency spectrum (collagen fibers) and areas of low image intensity (cytoplasm), 3D dynamic images with only nuclear component can be obtained in freshly excised tissues for the first time, as shown in Fig. 2C. Therefore, much more internal dynamic structures of the cell nuclei can be revealed. Benefiting from the reflectivity and dynamics information provided by active phase modulation, we can understand and infer these dynamic structures from multiple dimensions. As shown in Fig. 2J and 2K, it is interesting that the brightness, reflectivity, and dynamic intensity of these structures, drawn from 36 tumor cells in the sample, can be mainly categorized into four types, as indicated by black, blue, green arrows and purple dashed boxes in Fig.2e-h. The black arrows in Fig. 2F highlight rounded and black areas, and their 3D representations in Fig. 2G display dark spherical cavities. These structures closely resemble the spherical nucleoli seen in the H&E-stained images from the same sample, as shown in Fig. 2H. According to the literature, the nucleolus exhibits a relatively solid and compact nature within the nucleus[34], leading to higher reflectivity[35] and reduced dynamics[36] compared to the surrounding nucleoplasm. These properties are very consistent with the reflectivity and dynamics obtained by APMD-FFOCT, as shown in Fig. 2J, which further confirms that the black rounded areas correspond to nucleoli. As we know the size and activity of nucleoli exhibit a strong correlation with the tumor's malignancy(*37-39*), therefore this novel technique enables quick and label-free diagnosis of tumors based on nucleoli imaging. In Fig. 2F, the red arrows point to the black lines in tumor nuclei, while the 3D image in Fig. 2G reveals that these structures separate the nuclei completely. Thus, the red arrows mark the overlapping nuclear envelopes of adjacent nuclei in multinucleated cells, indicating a high degree of nuclear pleomorphism(*40*). In addition, there are some irregular dark blue structures distributed within nuclei, as indicated by the blue arrows. As shown in Fig. 2J, these structures exhibit the lowest reflectivity and highest dynamics, implying that they are loosely packed or dilute. The rest green parts of nuclei, primarily localized at the nuclear periphery, display moderate dynamics and reflectivity. To better understand the nature of the dark blue and green structures, further information is required.

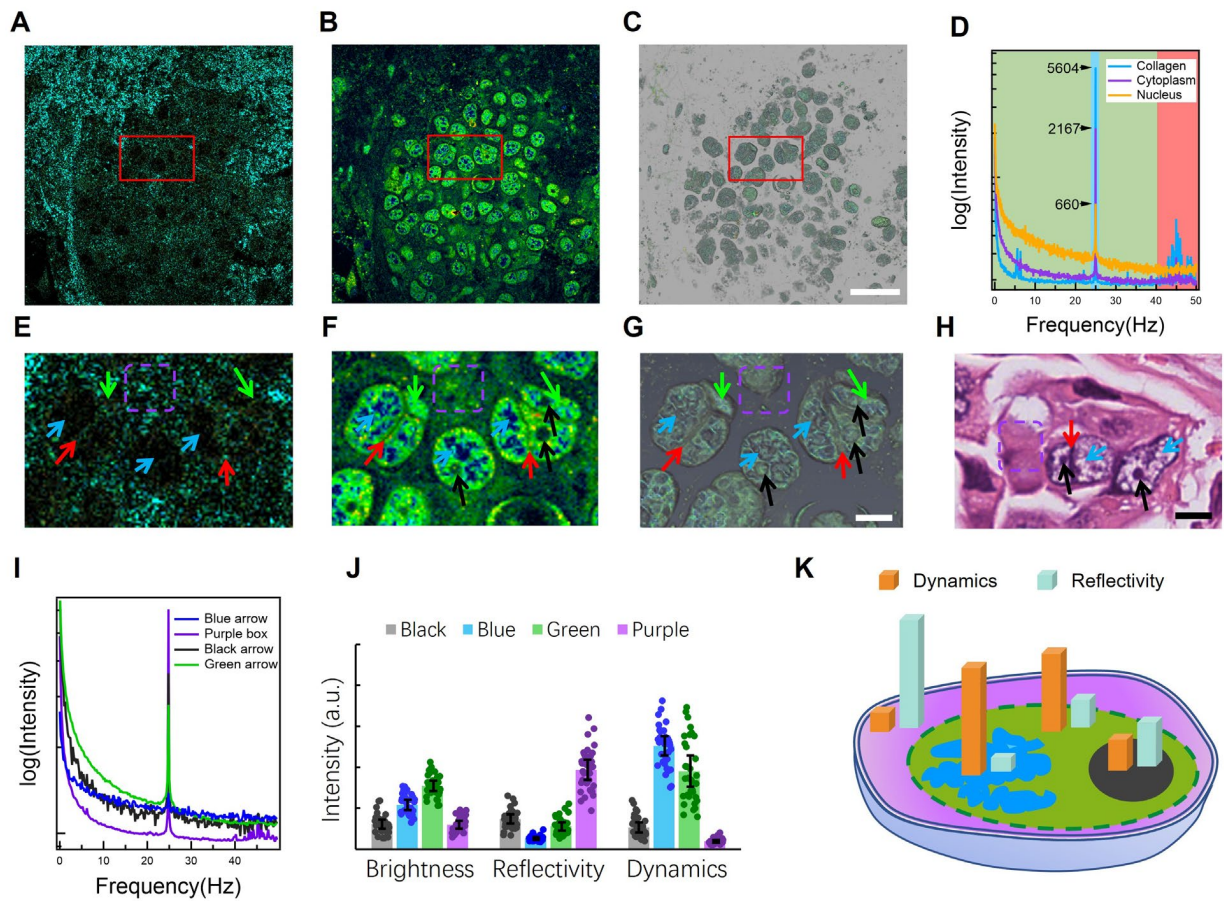

**Fig. 2. APMD-FFOCT imaging analysis of IDC samples: decoupling metabolic dynamics from reflectivity.** (**A**) Reflectivity image generated by active phase modulation part (blue area in (D)) of the frequency spectrum. (**B**) D-FFOCT image generated by frequency spectrum in the green area of (D). (**C**) Three-dimensional image of the tumor cell nuclei in tissue. (**D**) The frequency spectra of collagen fibers, nuclei, and cytoplasm in (A-C). (**E-G**) Enlargement of the red boxes in (A–C), highlighting specific features within the cells. Blue arrows point to dark blue areas in nuclei, red arrows point to the nuclear envelopes, black arrows point to black rounded areas, green arrows point to the rest part of nuclei, and the purple dotted box indicates cytoplasm. (**H**) Corresponding H&E histology image from the same sample. (**I**) Frequency spectra corresponding to the structures pointed by the blue arrows, black arrows, green arrows, and the cytoplasm in the purple dashed box in (E-G). (**J**) Statistical analysis of brightness, reflectivity, and dynamic intensity corresponding to structures in (E-G), N=36. The bars correspond to structures with markers in the same color in (E-G). Error bars: ±1 SD (standard deviation) Dynamics is obtained by dividing brightness

(frequency spectrum integral) by reflectivity (height of the active phase modulation peak). (**K**) Diagram showing the distribution of reflectivity and metabolic dynamics in tumor cells. Scale bars, 50 μm for (A-C), 10 μm for (E-H).

As a label-free imaging modality, APMD-FFOCT utilizes low-power illumination and hence allows for the long-term observation of tumor cells. It is therefore suitable to investigate the dynamic processes in freshly excised tissues, such as apoptosis. These dynamic processes can further provide additional information to identify the dark blue structures. Extended observations over an hour showed that the dark blue structures initially appeared at the center of the nuclei and gradually expanded towards the periphery, as demonstrated in Fig. 3A-C. During this period, the proportion of dark blue areas to the overall nuclear areas steadily increased, as depicted in Fig. 3D. This indicates that the dark blue structures are related to a certain process within the cells. Another observation spanning 12 hours in Fig. 3E-G revealed the distinct fate of these tumor cells. Specifically, Cell 4 (marked by red boxes) maintained its structural integrity throughout the observation. During this time, the brightness of Cell 4 gradually decreased, indicating a decline in metabolic dynamics, which may be due to gradual energy depletion. In contrast, Cell 5 (marked by orange boxes) exhibited a dark blue structure in the center of the nucleus and this structure gradually expanded to the periphery. Finally, the cell entered into a disordered state with low brightness, indicating cell death. Cell 6 in the blue boxes initially exhibited a uniform nucleus, after 3 hours, a dark blue structure appeared at the center of the nucleus. Eventually, this tumor cell exhibited dark and blurry structures, also indicating cell death. In fact, these processes are highly consistent with apoptosis, where chromatin condenses at the nuclear periphery, and then the nucleus fragments in several membrane-bound vesicles(*41*). Therefore, the dark blue structures likely represent residual nucleoplasm left after peripheral chromatin condensation, consistent with their low reflectivity and high dynamics characteristics. Moreover, Fig. 3H and Fig. 3I quantitatively depict the changes in mean dynamic intensity and reflectivity of Cells 4-6 over time. As shown in Fig. 3H. the dynamic intensity of Cell 4 initially displayed a plateau phase, lasting approximately 4 hours, then slowly and persistently declined. Meanwhile, the reflectivity of Cell 4 in Fig. 3I exhibited no obvious tendency until 10 hours and then slightly increased, indicating no obvious change in its structure. Conversely, Cells 5 and 6 exhibited a rapid reduction in dynamic intensity within the first 6 hours, subsequently stabilizing at low-intensity plateaus. Particularly, despite the later emergence of the dark blue structure in Cell 6, its dynamic intensity showed a more rapid decline. As the dynamic intensity of Cells 5 and 6 decreased, their reflectivity, in contrast, experienced an increase. Specifically, Cell 5's reflectivity increased for around 10 hours before stabilizing into a plateau phase, while Cell 6's reflectivity remained stable during the initial 3 hours, but following the emergence of the dark blue structure, its reflectivity rapidly increased and then transitioned to a phase of gradual ascent. We suggest that the rapid decline of dynamic intensity corresponds to reduced metabolic activity during the apoptosis, while the increase in reflectivity aligns with the chromatin condensation at the nuclear periphery and subsequent nuclear fragmentation into membrane-bound vesicles. Because that the transition from a loose and uniform state to a more compact and discontinuous state increases the refractive index difference, leading to an increased reflectivity. Therefore, the appearance of dark blue structures may be an indicator of apoptosis. And the apoptosis of tumor cells in tissues can be continuously monitored without the need for labeling or the risk of phototoxicity. Finally, the remaining green portion should be the nucleoplasm, packed with chromatin.

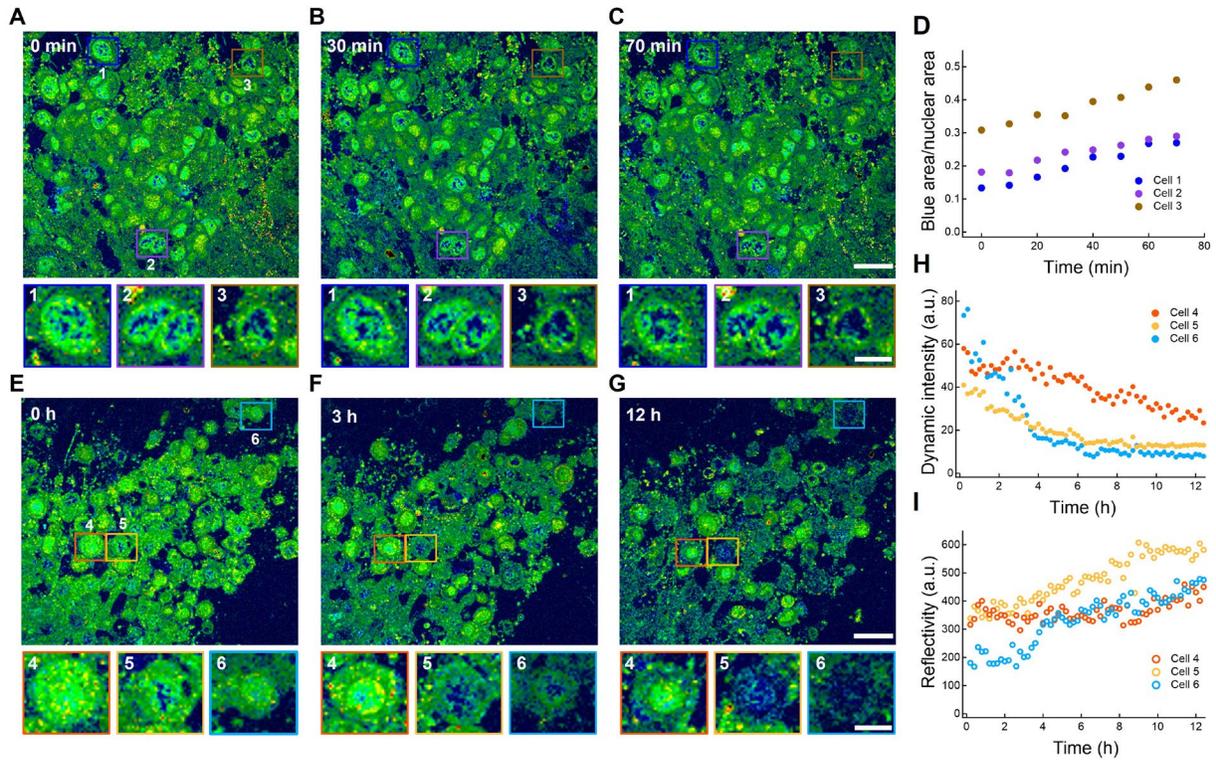

**Fig.3**. **Time-lapse images of tumor cells over periods of 1 hour and 12 hours.** (**A-C**) APMD-FFOCT images of the breast tumor captured at 0 minutes, 30 minutes, and 70 minutes, with insets showing magnified views of selected cells in boxes. (**D**) Ratios of the dark blue areas to the total areas of the cell nuclei change over time, corresponding to selected cells in (A-C). (**E-G**) APMD-FFOCT images of the breast tumor captured at 0 hours, 3 hours, and 12 hours, with insets displaying magnified views of the cells highlighted in the boxes. (**H-I**) Temporal variations of the dynamic intensity and reflectivity of the cells highlighted in the insets from (E-G), respectively. Scale bars, 30 μm for (A-C) and (E-G), 10 μm for the enlargements of selected cells.

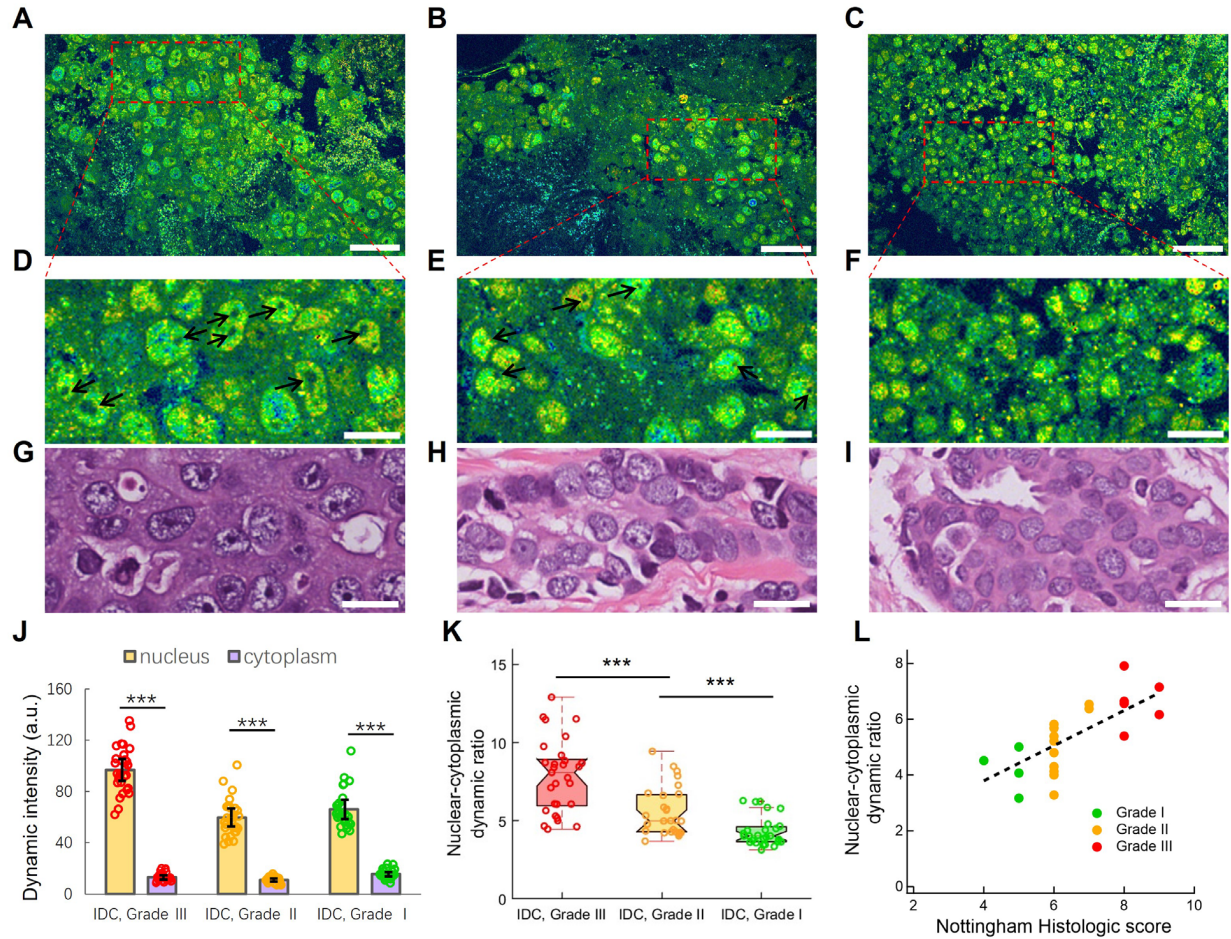

**Fig. 4. APMD-FFOCT images of different IDC grade and their dynamic intensity ratio of nucleus to cytoplasm. (A-C)** APMD-FFOCT images of IDC samples of grade III, grade II, and grade I, respectively. **(D-F)** Enlargement of the red dotted boxes in (A-C). **(G-I)** H&E-stained images from the sample corresponding to (A-C). **(J)** Statistical results of nuclear and cytoplasm dynamic intensity from the sample corresponding to (A-C), each point represents the averaged dynamic intensity of nuclei or cytoplasm from a subgraph in a stitched image. Error bars: ±1 SD. **(K)** Nuclear-cytoplasmic dynamic ratio obtained by dividing nuclear dynamic intensity by cytoplasm dynamic intensity in (J). **(L)** Nuclear-cytoplasmic dynamic ratio for increasing Nottingham Histologic score, each point represents the mean value of an independent sample. ***P < 0.0001. Scale bars, 50 μm for (A-C), 20 μm for (D-I).

### Dynamic intensity ratio of nucleus to cytoplasm as a biomarker of tumor malignancy

As previously mentioned, APMD-FFOCT has the ability to provide indicators of biological processes from a metabolic perspective, thus inspiring us to explore more biomarkers of metabolic dynamics for predicting clinical tumor grading. APMD-FFOCT images from IDC samples of grade III, grade II, and grade I have been obtained as shown in Fig. 4A-C. The histological tumor grades of the samples are determined based on the Nottingham Histologic scores derived from their H&E-stained images, which are regarded as the gold standard. Our analysis starts with the morphology and structures of tumor cells. As depicted in Fig. 4D, the APMD-FFOCT image of grade III IDC exhibits a large nuclear area (mean area: 62 μm²) with more prominent nuclear pleomorphism. Additionally, the tumor cells exhibit more and larger nucleoli, which indicates the lower differentiation and more malignant behavior of the sample(*37*). The corresponding H&E-stained image in Fig. 4G confirms our observation. Fig. 4E from grade II IDC shows a moderate nuclear size (mean area: 45 $\mu m^2$) combined with fewer and smaller nucleoli, which implies a lower malignant degree. The corresponding H&E-stained image in Fig. 4H shows similar structures. Fig. 4F from grade I IDC shows the smallest nuclear size (mean area: 38 $\mu m^2$) and there are almost no visible nucleoli, which is consistent with the histological image in Fig. 4I. Therefore, despite originating from different contrasts, APMD-FFOCT images also provide structural details consistent with H&E-stained images. Advancing further, we extracted dynamic intensity from the nuclei and cytoplasm of these three samples (see Supplementary Methods and fig. S6). As shown in Fig. 4J, the dynamic intensity of nuclei in three samples is significantly higher than their cytoplasm. Specifically, the nuclear dynamic intensity of the grade III sample is higher than that of other grades, while the cytoplasmic dynamic intensity is highest in the grade I sample. However, the basal metabolic rate, which affects the metabolic activities of both the nucleus and the cytoplasm, varies across different samples due to individual patient factors, including genetic makeup(42), age(43), gender(44), hormonal profiles(45), and lifestyle. To further reduce the impact of different basal metabolic rates, we extract the nuclear-cytoplasmic dynamic ratio (NCDR) by dividing nuclear dynamic intensity by their cytoplasmic dynamic intensity, as shown in Fig. 4K. The result shows a positive correlation between the NCDR and tumor grading. We then conducted additional experiments on 21 IDC samples, each from an independent patient (Supplementary Table S1). As shown in Fig. 4L, each point represents the averaged NCDR for an individual sample. The statistical result shows a positive correlation between NCDR and the Nottingham Histologic score of the samples (r = 0.74, P = 1.1e-4), where "r" represents the Pearson correlation coefficient, and "P" signifies the significance level of the correlation. Individually, scatter plots (fig. S7 B-C) show an upward trend of nuclear dynamic intensity (r = 0.31, P = 0.18) and a downward trend of the cytoplasmic dynamic intensity (r = -0.53, P = 0.013) with increasing Nottingham Histologic score. These findings prove that NCDR exhibits a more pronounced correlation with tumor grading, which indicates that NCDR could serve as an effective biomarker for predicting tumor malignancy. In addition, we also explore the relationship between NCDR and nuclear area (fig. S7d), because nuclear area is also a parameter of tumor grading. The result shows a more obvious correlation with prominent significance (r = 0.81, P = 7e-6). This is because the NCDR and nuclear area are from the same APMD-FFOCT images and hence there exists no decorrelation due to the tumor heterogeneity and the sampling misalignment between H&E-stained images and APMD-FFOCT images. It is worth pointing out that this correlation tends to be non-linear, as shown in fig. S7d. Individually, the nuclear dynamic intensity increases with the increasing of nuclear area (r = 0.61, P = 0.003), while the cytoplasmic dynamic intensity shows the opposite trend (r = -0.24, P = 0.29), as shown in fig. S7E-F. Therefore, the dependence of NCDR on the nuclear area demonstrates a significantly stronger correlation and reduced variance compared to the dependence on nuclear dynamics or cytoplasmic dynamics alone, which further confirms our assumption that NCDR reduces the effects of basic metabolism differences.

## Discussion

We have developed an APMD-FFOCT for imaging and metabolic analysis of freshly excised specimens without any processing. By treating the active modulation as a reference movement for all interferential scatterers, the metabolic dynamics of these scatterers can be decoupled from their reflectivity and other imaging factors. Within one area, factors like illumination and interference are uniform, so the relative reflectivity can be approximated as the height of the active phase modulation peaks in the frequency spectra. Although the reflectivity of tissues can also be obtained by FFOCT imaging(24), there are shortcomings in this method. Firstly, implementing FFOCT imaging means additional time is required, which is unacceptable during surgery. And because the imaging time is separated and the scatterers keep moving, the scatterers measured by FFOCT may not be consistent with the scatterers measured by APMD-FFOCT. Secondly, the response sensitivity of interferometry is related to the interference phase of the scatterers, as the active phase modulation is a small amplitude oscillation near the phase where the scatterers are located. Therefore, it can reflect the impact of phase factors in interferometric measurements, resulting in a more accurate separation and quantification of metabolic dynamics (fig. S8).

Despite coming from different sources of contrast, APMD-FFOCT provides comparable structural details to H&E-stained images in 3D with additional dynamics and reflectivity information (Fig. 2). Nuclei, cytoplasm, and collagen fibers in the same image can be distinguished according to their difference in dynamics and reflectivity. Combined with low-coherence gating, high-resolution 3D nuclear contours and structures can be obtained without labeling. As an important index of the Nottingham Grading System, nuclear pleomorphism is one of the least reproducible parameters to score histological grade, due to the 2D analyses of histological specimens. Shifting from 2D to 3D imagery could improve the accuracy of clinical diagnosis(*46*), yet traditional 3D nuclear structure imaging methods, involving complex fluorescent labeling, limit their clinical use(*47, 48*). Therefore, the convenient and rapid 3D tomography capability of APMD-FFOCT offers a comprehensive and intuitive representation of nuclear pleomorphism, which could enhance the accuracy of pathological diagnoses. Moreover, combined with the dynamics and reflectivity analysis, subnuclear structures including nucleolus, chromatin, and the nuclear envelope can be imaged and deduced, which are indispensable for pathological diagnosis. For example, the hypertrophy of the nucleolus is one of the most distinctive cytological features of cancer cells as well as important parameters to score histological grade. It has proved that nucleolar area values were inversely related to the tumor mass double time(*37, 38*). Therefore, APMD-FFOCT represents a significant advancement in label-free 3D imaging for pathological analysis, providing detailed structural and dynamic information that may enhance diagnostic accuracy and clinical utility.

Apoptosis is a key cell death mechanism in tumors responding to treatment(*49*). Changes in the tumor microenvironment, such as insufficient supply of nutrients and oxygen, can also induce apoptosis(*50*). Extended observations of tumor cells over time reveal a process highly consistent with apoptosis (Fig. 3), where chromatin gradually condensed at the nuclear periphery, followed by the collapse of the cell structures(*41*). The changes in dynamics and reflectivity over time, extracted from the cells, further confirm this apoptotic process. We show that APMD-FFOCT provides a unique perspective to monitor the apoptosis of tumor cells in tissues without the need for labeling or the risk of photodamage. This capability holds substantial significance for evaluating the effectiveness of chemotherapy and various drugs in inducing tumor apoptosis(*49*).

Furthermore, the decoupling of dynamic intensity enables the horizontal comparison of different samples or different areas within one sample, thereby holding the promise of providing a biomarker of tumor malignancy based on subcellular metabolism. Dynamic images from IDC samples of grade III, grade II, and grade I correspond well with the H&E-stained images (Fig. 4). To further minimize the effects of basal metabolic rate variations caused by individual factors or environmental conditions, we came up with NCDR as a biomarker, which shows a positive correlation with tumor grading. Experiments on 21 samples from independent patients provided additional evidence for this perspective with high significance. Individually, scatter plots show an upward trend of nuclear dynamic intensity and a downward trend of the cytoplasmic dynamic intensity with increasing Nottingham Histologic score. One explanation for the observed phenomena is that as tumor cell differentiation decreases (indicative of a high-grade tumor), their cytoplasmic functions deteriorate, leading to a decline in cytoplasmic metabolism. Concurrently, as the malignancy level increases, the tumor proliferation rate rises, leading to an increase in nuclear dynamics. And the excessively high metabolic demands of the cell nucleus deplete the energy resources of the cytoplasm, which further reduces the metabolic dynamics of the cytoplasm. One evidence for this explanation is that, as a type of highly differentiated functional cell, liver cells hardly proliferate and undertake important metabolic activities, the dynamic intensity of their nuclei is lower than that of their cytoplasm, in contrast to tumor cells, as shown in fig. S9.

Extracting metabolic dynamics as an objective biomarker through optical methods faces numerous challenges. Firstly, the measurement of metabolic dynamics is influenced by various factors, such as the reflectivity of scatterers, the illumination intensity on the focal plane, the light attenuation in the optical path, and the defocus effect around a depth of interest. By employing active modulation to provide a standard reference motion, we can decouple these factors from metabolic dynamics. Furthermore, the basal metabolic rate of biological tissues is also affected by multiple factors, including individual differences and environmental factors. By comparing metabolic dynamics between the nucleus and cytoplasm, we can further reduce the impact of these factors. After two rounds of comparison, we obtained NCDR as a relatively objective biomarker that reflects the characteristics of tumor metabolism and further demonstrated its association with the malignancy degree of IDC. NCDR is expected to optically provide a biomarker of tumor malignancy based on metabolic dynamics in addition to pathological morphology and molecular markers. This metabolic biomarker can be acquired at the same time as obtaining pathological images and does not require any additional tissue processing, which would benefit the rapid and automated intraoperative diagnosis.

**Materials and methods**

**Experimental setup**

The APMD-FFOCT configuration utilizes a Linnik interferometer with twin 20x water immersion objectives and a low-coherence light source. Light from LED (M565L3, Thorlabs, center wavelength of 565 nm and a full width at half maximum of 104 nm) was split into the sample arm and reference arm using a 50/50 nonpolarized beam splitter (BS013, Thorlabs), and then focused at the rear focal plane of objectives to ensure Köhler illumination. We use a pair of identical water immersion objectives (Nikon NIR APO 20x 0.5 NA) to achieve ~0.7 μm lateral resolution with a field of view of ~500 × 350 μm². The light power irradiated on the sample is 1.2 mW. Using water as a medium allows for refractive index matching to reduce surface reflections from the objective lens and cover slides. Simultaneously, it minimizes the separation between the focal plane and the coherence plane when adjusting imaging depth(*28*). The axial resolution is ~1 μm decided by the coherence length of the light source. In the sample arm, the freshly excised tissues are mounted on a customed sample holder. Tissues were kept moist with saline and gently pressed down with a coverslip to create a flatter imaging surface. The height of the coverslip can be adjusted by rotating the thread and is locked by a clasp to ensure that the position of the cover glass does not move. Then the sample holder was mounted on a five-dimensional control table with three dimensions for electric translation and two dimensions for Angle adjustment, which ensures the cover glass is parallel to the focal surface, so the imaging depth is consistent when the translation table is transversally translated. In the reference arm, the reference mirror (YAG, Yttrium Aluminum Garnet) can be modulated by the PZT (TA0505D024W, Thorlabs) beneath it, which can be used to produce phase modulation when imaging. The whole reference arm is mounted on a high-precision electric translation platform (M-VP-25XL, Newport) to adjust the light path difference between the two arms. The focal plane of objectives should overlap with the coherence plane to achieve the best performance of imaging(*28*). Back-scattered light from subcellular scatterers and reference mirror overlap after passing through the beam splitter again. The combined light is then focused on the camera (MV4-D1600-S01-GT, Photon Focus) by a tube lens. The whole setup was mounted on an active vibration isolation platform (VCM-S400) to reduce the environmental vibration.

**Data acquisition, processing, and image generation**

In our study, all APMD-FFOCT control and raw image acquisition were executed by a custom C++ based software. The acquisition time of a dynamic image is set at 5 seconds, with a frame rate of 100 FPS, to capture a wide range of intracellular movements. For each acquisition, we typically obtain a (550, 800, 500) tensor in which 550 × 800 is the number of sensor pixels after pixel binning (2,2) and 500 is the number of recorded frames. Binning reduces the number of pixels and therefore saves storage space. Meanwhile, it increases the quantum well depth by four times, which improves the signal-to-noise ratio. Each pixel generates a time-domain intensity trace with 500 points over a 5-second interval, which is subsequently Fourier-transformed to obtain the frequency spectrum. The lowest detectable motion frequency (0.2 Hz) is determined by the total acquisition time, while the highest detectable motion frequency (50 Hz) is determined by the camera frame rate. We then employed the Hue-Saturation-Value (HSV) color space to visualize the dynamic characteristics of the metabolism activities. Hue is determined by the spectral centroid.

$$\text{Hue} = \frac{\sum_{n=1}^{N} f(n) \cdot S(f(n))}{\sum_{n=1}^{N} S(f(n))} \qquad (1)$$

$N$ is the number of points in the spectrum, $f(n)$ is the value of frequency, representing each frequency point in a discrete Fourier transform. $S(f(n))$ is the spectral intensity at the frequency f(n). This formula calculates the weighted average of each frequency point in the spectrum. Therefore, hue represents the average fluctuation speed. The brightness of the image is determined by the integration of the frequency spectrum, excluding the zero-frequency portion (represents static background), which represents fluctuation amplitude.

$$Brightness = \sum_{n=1}^{N} S(f(n)) \qquad (2)$$

Finally, saturation was computed as the standard deviation of the frequencies. To some extent, saturation reflects the complexity of motions. Saturation is quantified as follows, where $sc$ denotes the spectral centroid:

$$\text{Saturation} = \sqrt{\frac{\sum_{n=1}^{N} (f(n) - sc)^2 \cdot S(f(n))}{\sum_{n=1}^{N} S(f(n))}} \qquad (3)$$

Then, the above parameters are normalized by a linear transformation. Finally, the HSV image is transformed in the RGB color space to display. All data processing and analysis are done on our custom software based on MATLAB.

**Principle of APMD-FFOCT**

The reflectivity of the scatterers located in the coherence slice of the sample at a given depth is $R(x, y)$. The rest of the light backscattered from the sample and collected by the microscope objective that does not interfere is represented by an equivalent reflectivity coefficient denoted as $R_{Inc}$. The reflectivity of the reference mirror is uniform, represented by $R_{ref}$. The axial position of scatterers and reference mirror are $P_s(x, y, t)$ and $P_r(t)$, respectively. As shown in the formula, the modulation of interference light intensity depends not only on the motion of scatterers but also on the reflectivity of the scatterers. The intensity of light incident onto the CMOS can be written as:

$$I(x,y,t) = \frac{I_0(x,y)}{4} \cdot \left( R_{Inc} + R_{ref} + R(x,y) + 2\sqrt{R(x,y) \cdot R_{ref}} \cdot \cos\left( 2\pi \cdot \frac{P_s(x,y,t) - P_r(t)}{\lambda} \right) \right) \quad (4)$$

where $I_0$ is the intensity of light incident on the interferometer beam splitter. The first three items in the formula contribute to a static background and can therefore be ignored. We applied a sinusoidal voltage with a frequency of 25 Hz and an amplitude of 0.5 V to the PZT to generate a sinusoidal motion in the reference mirror with a frequency of 25 Hz and an amplitude of ~23 nm.

$$P_r(t) = A_{pzt} \cdot e^{i2\pi \cdot f_0 \cdot t} \quad (5)$$

The metabolic motion of subcellular scatterers can be decomposed into the superposition of multiple sinusoidal motions at different frequencies.

$$P_s(x,y,t) = \int_0^\infty A_s(x,y,f) \cdot e^{i2\pi \cdot f \cdot t} df \quad (6)$$

Therefore, the formula can be written as:

$$I(x,y,t) = \frac{I_0(x,y)}{2} \cdot \sqrt{R(x,y) \cdot R_{ref}} \cdot \cos\left( 2\pi \cdot \frac{\int_0^\infty A_s(x,y,f) \cdot e^{i2\pi \cdot f \cdot t} df - A_{pzt} \cdot e^{i2\pi \cdot f_0 \cdot t}}{\lambda} \right) + BG \quad (7)$$

The $BG$ represents the part of light intensity that does not change with time. To simplify the derivation, we make a linear approximation of the cosine term and ignore the constant.

$$I(x,y,t) = \frac{I_0(x,y)}{2} \cdot \sqrt{R(x,y) \cdot R_{ref}} \cdot \left( 2\pi \cdot \frac{\int_0^\infty A_s(x,y,f) \cdot e^{i2\pi \cdot f \cdot t} df - A_{pzt} \cdot e^{i2\pi \cdot f_0 \cdot t}}{\lambda} \right) \quad (8)$$

After the Fourier transform, each frequency component will form the corresponding spectral intensity. The active phase modulation forms a sharp intensity peak at 25Hz. The spectrum can be approximated as:

$$\mathcal{F}\{I(x,y,t)\} = \frac{\pi I_0(x,y)}{\lambda} \cdot \sqrt{R(x,y) \cdot R_{ref}} \cdot \left( A_s(x,y,f) - A_{pzt}\delta(f - f_0) \right) \quad (9)$$

This result includes two main frequency components: the low-frequency part $A_s(x, y, f)$ reflecting the metabolic motion of the scatterers and the high-frequency part $A_{pzt}\delta(f - f_0)$ reflecting the active modulation by the PZT. Therefore, metabolic dynamics can be decoupled from reflectivity and incident light distribution by dividing the spectral integral term by the active modulation peak term.

$$Dynamic\ intensity = \int_{f=0.4}^{f=50} A_s(x,y,f) / A_{pzt}(f_0) \quad (10)$$

If we limit the spectrum to 25Hz, which means that $A_s(x, y, f)$ can be ignored. Considering that Kohler lighting is uniformly illuminated in the field of view, the generated image approximately reflects the reflectivity distribution.

$$Reflectivity = \frac{I_0}{2} \cdot \sqrt{R(x,y) \cdot R_{ref}} A_{pzt}(f_0) \qquad (11)$$

**Data analysis**

Data analysis is provided in Supplementary Methods.

**Sample**

All study procedures were approved by the Ethics Committee of Peking University People's Hospital. Informed consent was obtained from each patient undergoing breast cancer surgery before imaging. Fresh breast tumor tissue was promptly collected from the tumor bed of the excised surgical specimen, ensuring the presence of at least one smooth and flat surface suitable for APMD-FFOCT imaging. The excised tissue was kept moist with saline and the container was stored on ice until imaging. All data in Fig. 4 were obtained at room temperature (~25 °C) within 2 hours after excision, during which period the changes in cellular metabolic dynamics were minimal (fig. S10). After APMD-FFOCT imaging, the tissue was fixed in formalin for paraffin H&E pathology. A pathologist with expertise in breast histology provided a pathological diagnosis for each tissue based on the corresponding H&E slide. Tissues diagnosed as non-IDC or ductal carcinoma in situ (DCIS) were excluded from this study. Then, the experienced pathologist assessed the histological grade of IDC slides according to the Nottingham grading system.

**References and Notes**

1.    N. N. Pavlova, C. B. Thompson, The Emerging Hallmarks of Cancer Metabolism. *Cell Metab* **23**, 27-47 (2016).
2.    D. Hanahan, R. A. Weinberg, Hallmarks of cancer: the next generation. *Cell* **144**, 646-674 (2011).
3.    R. A. Gatenby, R. J. Gillies, Why do cancers have high aerobic glycolysis? *Nat Rev Cancer* **4**, 891-899 (2004).
4.    U. E. Martinez-Outschoorn, M. Peiris-Pages, R. G. Pestell, F. Sotgia, M. P. Lisanti, Cancer metabolism: a therapeutic perspective. *Nat Rev Clin Oncol* **14**, 11-31 (2017).
5.    S. Hakomori, Tumor malignancy defined by aberrant glycosylation and sphingo(glyco)lipid metabolism. *Cancer Res* **56**, 5309-5318 (1996).
6.    M. Titford, A short history of histopathology technique. *Journal of Histotechnology* **29**, 99-110 (2006).
7.    R. J. Buesa, Productivity standards for histology laboratories. *Ann Diagn Pathol* **14**, 107-124 (2010).
8.    T. A. Holly *et al.*, Single photon-emission computed tomography. *J Nucl Cardiol* **17**, 941-973 (2010).
9.    M. D. Farwell, D. A. Pryma, D. A. Mankoff, PET/CT imaging in cancer: Current applications and future directions. *Cancer* **120**, 3433-3445 (2014).
10.   D. E. Befroy, G. I. Shulman, Magnetic resonance spectroscopy studies of human metabolism. *Diabetes* **60**, 1361-1369 (2011).
11.   M. Momcilovic, D. B. Shackelford, Imaging Cancer Metabolism. *Biomol Ther (Seoul)* **26**, 81-92 (2018).

12.	A. R. Pantel, D. Ackerman, S. C. Lee, D. A. Mankoff, T. P. Gade, Imaging Cancer Metabolism: Underlying Biology and Emerging Strategies. *J Nucl Med* **59**, 1340-1349 (2018).

13.	I. S. Gilmore, S. Heiles, C. L. Pieterse, Metabolic Imaging at the Single-Cell Scale: Recent Advances in Mass Spectrometry Imaging. *Annu Rev Anal Chem (Palo Alto Calif)* **12**, 201-224 (2019).

14.	D. W. Shipp, F. Sinjab, I. Notingher, Raman spectroscopy: techniques and applications in the life sciences. *Advances in Optics and Photonics* **9**, 315-428 (2017).

15.	C. J. Sheppard, Multiphoton microscopy: a personal historical review, with some future predictions. *Journal of biomedical optics* **25**, 014511-014511 (2020).

16.	V. Ruiz-Rodado, A. Lita, M. Larion, Advances in measuring cancer cell metabolism with subcellular resolution. *Nature Methods* **19**, 1048-1063 (2022).

17.	M. C. Skala *et al.*, In vivo multiphoton microscopy of NADH and FAD redox states, fluorescence lifetimes, and cellular morphology in precancerous epithelia. *Proceedings of the National Academy of Sciences* **104**, 19494-19499 (2007).

18.	Z. Liu, L. D. Lavis, E. Betzig, Imaging live-cell dynamics and structure at the single-molecule level. *Mol Cell* **58**, 644-659 (2015).

19.	C. P. Brangwynne, G. H. Koenderink, F. C. MacKintosh, D. A. Weitz, Intracellular transport by active diffusion. *Trends Cell Biol* **19**, 423-427 (2009).

20.	G. Salbreux, G. Charras, E. Paluch, Actin cortex mechanics and cellular morphogenesis. *Trends Cell Biol* **22**, 536-545 (2012).

21.	V. Ntziachristos, Going deeper than microscopy: the optical imaging frontier in biology. *Nat Methods* **7**, 603-614 (2010).

22.	C. Apelian, F. Harms, O. Thouvenin, A. C. Boccara, Dynamic full field optical coherence tomography: subcellular metabolic contrast revealed in tissues by interferometric signals temporal analysis. *Biomed Opt Express* **7**, 1511-1524 (2016).

23.	J. Scholler *et al.*, Dynamic full-field optical coherence tomography: 3D live-imaging of retinal organoids. *Light Sci Appl* **9**, 140 (2020).

24.	A. Dubois, L. Vabre, A. C. Boccara, E. Beaurepaire, High-resolution full-field optical coherence tomography with a Linnik microscope. *Appl Opt* **41**, 805-812 (2002).

25.	H. Yang *et al.*, Use of high-resolution full-field optical coherence tomography and dynamic cell imaging for rapid intraoperative diagnosis during breast cancer surgery. *Cancer* **126 Suppl 16**, 3847-3856 (2020).

26.	Ultrahigh-resolution full-field optical coherence tomography.

27.	S. Zhang *et al.*, Potential rapid intraoperative cancer diagnosis using dynamic full-field optical coherence tomography and deep learning: A prospective cohort study in breast cancer patients. *Science Bulletin*, (2024).

28.	S. Labiau, G. David, S. Gigan, A. C. Boccara, Defocus test and defocus correction in full-field optical coherence tomography. *Opt Lett* **34**, 1576-1578 (2009).

29.	D. D. Nolte, R. An, J. Turek, K. Jeong, Holographic tissue dynamics spectroscopy. *J Biomed Opt* **16**, 087004 (2011).

30.	N. Monnier *et al.*, Inferring transient particle transport dynamics in live cells. *Nat Methods* **12**, 838-840 (2015).

31.	M. Guo *et al.*, Probing the stochastic, motor-driven properties of the cytoplasm using force spectrum microscopy. *Cell* **158**, 822-832 (2014).

32.	W. Han *et al.*, Oriented collagen fibers direct tumor cell intravasation. *Proc Natl Acad Sci U S A* **113**, 11208-11213 (2016).

33.	M. W. Conklin *et al.*, Aligned Collagen Is a Prognostic Signature for Survival in Human Breast Carcinoma. *The American Journal of Pathology* **178**, 1221-1232 (2011).

34. T. Pederson, The nucleolus. *Cold Spring Harb Perspect Biol* **3**, (2011).
35. M. Fimiani, P. Rubegni, E. Cinotti, *Technology in Practical Dermatology: Non-Invasive Imaging, Lasers and Ulcer Management.* (Springer Nature, 2020).
36. A. Zidovska, The rich inner life of the cell nucleus: dynamic organization, active flows, and emergent rheology. *Biophysical Reviews* **12**, 1093-1106 (2020).
37. M. Derenzini *et al.*, Nucleolar size indicates the rapidity of cell proliferation in cancer tissues. *The Journal of Pathology* **191**, 181-186 (2000).
38. D. Treré, C. Ceccarelli, L. Montanaro, E. Tosti, M. Derenzini, Nucleolar Size and Activity Are Related to pRb and p53 Status in Human Breast Cancer. *Journal of Histochemistry & Cytochemistry* **52**, 1601-1607 (2016).
39. M. Derenzini *et al.*, Nucleolar function and size in cancer cells. *Am J Pathol* **152**, 1291-1297 (1998).
40. S. A. Silver, F. A. Tavassoli, Pleomorphic carcinoma of the breast: clinicopathological analysis of 26 cases of an unusual high-grade phenotype of ductal carcinoma. *Histopathology* **36**, 505-514 (2000).
41. A. Wyllie, G. Beattie, A. Hargreaves, Chromatin changes in apoptosis. *The Histochemical Journal* **13**, 681-692 (1981).
42. R. A. Cairns, I. S. Harris, T. W. Mak, Regulation of cancer cell metabolism. *Nat Rev Cancer* **11**, 85-95 (2011).
43. C. C. Benz, Impact of aging on the biology of breast cancer. *Crit Rev Oncol Hematol* **66**, 65-74 (2008).
44. S. Haupt, F. Caramia, S. L. Klein, J. B. Rubin, Y. Haupt, Sex disparities matter in cancer development and therapy. *Nat Rev Cancer* **21**, 393-407 (2021).
45. S. Satpathi, S. S. Gaurkar, A. Potdukhe, M. B. Wanjari, Unveiling the Role of Hormonal Imbalance in Breast Cancer Development: A Comprehensive Review. *Cureus* **15**, e41737 (2023).
46. J. Najbauer *et al.*, Isotropic 3D Nuclear Morphometry of Normal, Fibrocystic and Malignant Breast Epithelial Cells Reveals New Structural Alterations. *PLoS ONE* **7**, (2012).
47. G. Bussolati, C. Marchiò, L. Gaetano, R. Lupo, A. Sapino, Pleomorphism of the nuclear envelope in breast cancer: a new approach to an old problem. *Journal of Cellular and Molecular Medicine* **12**, 209-218 (2007).
48. S. Liu, D. L. Weaver, D. J. Taatjes, Three-dimensional reconstruction by confocal laser scanning microscopy in routine pathologic specimens of benign and malignant lesions of the human breast. *Histochem Cell Biol* **107**, 267-278 (1997).
49. M. Garcia-Barros *et al.*, Tumor response to radiotherapy regulated by endothelial cell apoptosis. *Science* **300**, 1155-1159 (2003).
50. C. D. Gregory, J. D. Pound, Microenvironmental influences of apoptosis in vivo and in vitro. *Apoptosis* **15**, 1029-1049 (2010).

**Acknowledgments**

**Funding:**

Natural Science Foundation of China with Grant No. 61575108

Natural Science Foundation of China with Grant No. 61905015

Natural Science Foundation of China with Grant No. 61975091

Natural Science Foundation of China with Grant No. 82372629

Natural Science Foundation of China with Grant No. 92059105

Bio-Brain+X' Advanced Imaging Instrument Development Seed Grant

Tsinghua Precision Medicine Foundation

Beijing Municipal Science and Technology Project (Z201100005520081).

**Author contributions:**

Conceptualization: ZCY, SWZ, PX, SW

Methodology: ZCY, SWZ

Investigation: ZCY, SWZ,  BH.

System construction: ZCY, PQY

Samples preparation: SWZ, HPY, YZY

Visualization: ZCY, ZYC, ZWH

Data collection: ZCY, YJS, RZX, CMW

Funding acquisition: PX, SW

Project administration: PX, SW

Supervision: PX, SW

Writing – original draft: ZCY, SWZ

Writing – review & editing: BH, PX, SW

**Competing interests:** The authors declare no competing interests

**Data and materials availability:** All data and code supporting the findings of this study are available within this paper. The raw data, due to their large file size, are available from the corresponding author upon request.

## Supplementary Materials

Supplementary Methods

Figure. S1 to S10

Table S1

**Supplementary Methods**

**Quantitative analysis of the brightness, reflectivity, and dynamic intensity of subnuclear structures**

The camera in APMD-FFOCT is almost shot-noise limited. When their pixel wells are close to saturation, the electrical noise is small compared to the shot noise, which is Poisson-distributed. For an average photon count $N$, the standard deviation of the shot noise is $\sqrt{N}$. This Poisson noise results in a flat background across the frequency spectrum, with its height dependent on light intensity. In all of our analyses, this background of the frequency spectrum is subtracted according to the mean light intensity of the corresponding area.

The brightness, reflectivity, and dynamic intensity in Fig.2 i-j are extracted from the mean spectrum in manually selected rectangular regions corresponding to the subnuclear structures. Brightness is quantified by the integration of the frequency spectrum. "Reflectivity" refers to the height of active modulation intensity peak at 25Hz. The dynamic intensity of subcellular metabolism can be quantified by dividing brightness (frequency spectrum integral) by reflectivity (height of the active phase modulation peak).

**Nuclear-cytoplasmic dynamic ratio calculation**

To efficiently process thousands of tumor cells for the nuclear-cytoplasmic dynamic ratio calculation (Fig. 4) and minimize subjective biases, we developed an automated program based on MATLAB for nuclei and cytoplasm labeling and dynamic intensity calculation. Key steps include:

**1. Collagen fiber intensity reduction**: Isolate the low-frequency spectrum part to diminish collagen fiber intensity, highlighting cellular structures.

**2. Labeling the nuclear region**: Convert the image to grayscale and perform image binarization for initial segmentation, separating nuclear components. Then utilize the "bwareaopen" function to filter out undersized white areas in the binary images. This binary image marks the pixels occupied by nuclei, called initial nuclear binary image. The nuclear binary image then executed the "imerode" function to shrink the nuclear region to avoid the influence of the nuclear envelope. This image is called the nuclear binary image, which marks the central part of the nuclear region.

**3. Labeling the cytoplasmic region**: First, perform a minor dilation based on the initial nuclear binary image to cover both the nuclear envelope and nuclear region (To avoid the influence of the nuclear envelope). Then, conduct the second dilation and intersect with the former image to get the cytoplasmic binary image, which marks the cytoplasmic region.

**4. Calculation**: The final labeling result is shown in the Fig. S6. The nuclear and cytoplasmic dynamic intensity refers to the mean dynamic intensity of the respective areas marked by nuclear binary image and cytoplasmic binary image. Finally, the nuclear-cytoplasmic dynamic ratio can be obtained by dividing nuclear dynamic intensity by cytoplasmic dynamic intensity.

**Nuclear number and the average nuclear size calculation**

**1, Applying distance transform to the binary image:** Based on the above obtained initial nuclear binary image, which marks the pixels occupied by nuclei, a distance transform of the binary image was conducted to give the distance transform image, where the value of each pixel is the distance between that pixel and the nearest non-zero pixel in the binary image.

**2, Image segmentation with watershed algorithm:** Segment the distance transform image based on the watershed algorithm, delineating boundaries between connected or overlapping nuclei.

**3, Connected components analysis:** Identifies connected components in the segmented image to analyze clusters of pixels identified as individual nuclei. It extracts and calculates geometric properties such as perimeter and area for each detected nucleus in the image using the connected component data and stores these values in arrays. Consequently, nuclear number and average area size can be calculated (Fig. 4).

### Intranuclear dark blue area size calculation

In the initial nuclear binary image obtained above, if dark blue structures are present within the nuclei, their corresponding pixels will not be marked, resulting in holes within the marked nuclear area. The "imfill" function in MATLAB is then used to fill these holes. Subsequently, this filled image intersects with the former image to generate a new binary image that marks the regions occupied by the dark blue structures. Finally, the area of the dark blue structures is quantified using connected components analysis (Fig. 3).

### Statistical test

For data adhering to a normal distribution, independent and paired sample t-tests were conducted using MATLAB. Specifically, the "ttest2" function was employed for independent samples (Fig. 4k), while the "ttest" function was used for paired samples (Fig. 4j). Furthermore, the correlation and its significance between two variables were assessed using the "corr" function in MATLAB (Fig. 4l, Fig.S7). Pearson's correlation coefficient was computed using [r, p] = corr(var1, var2, 'Type', 'Pearson'), where var1 and var2 are the data vectors of the two variables, "r" represents the Pearson correlation coefficient, and "p" signifies the significance level of the correlation. Datasets were considered non-significant (n.s.) if $p>0.01$. The following significance symbols have been used for the corresponding p-values: * $p<0.01$, ** $p<0.001$ and *** $p<0.0001$.

**Supplementary Figures S1-S10**

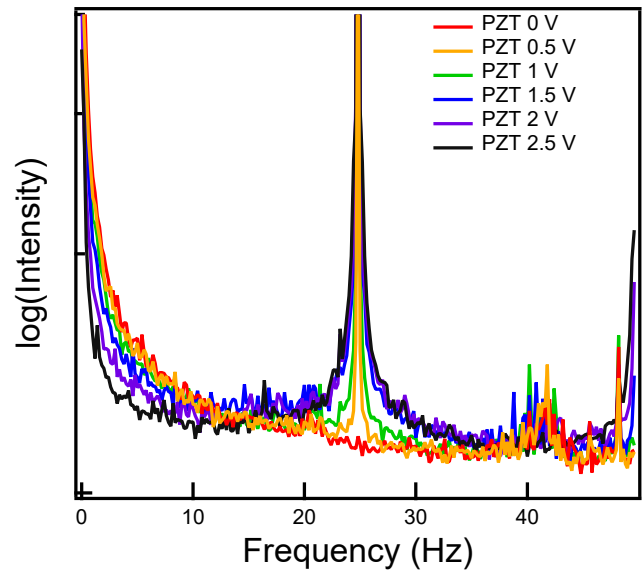

**Figure S1. Frequency spectra of tumor cells with increasing PZT modulation voltage.** With the increase of PZT modulation voltage, the active modulation peak gradually widens, and the spectra of metabolic activity are gradually suppressed. The relative difference between the spectrum integrals (from 0.4 Hz to 24 Hz) of PZT at 0 V and PZT at 0.5 V is approximately 1 %. The intensity peaks near 40 Hz are caused by ambient vibrations.

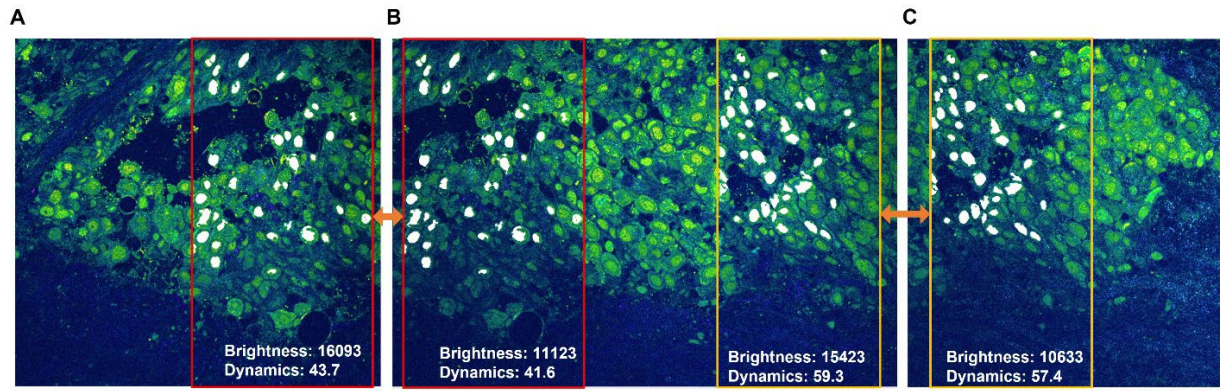

**Figure S2. Comparison of brightness and dynamic intensity of overlapping regions in continuous images.** (**A-C**) Three consecutive APMD-FFOCT images of the tumor cells in the IDC sample. The red box on the right in (A) and the red box on the left in (B) are overlapping areas, but they are in opposite relative positions in the field of view. The white area marks exactly the same nuclei used to calculate brightness and dynamics intensity. The illumination intensity and interference of the two red boxes are different, which also can be considered as different reflectivity of scatterers. The same process is in the orange boxes in (B) and (C). The relative brightness difference between the marked nuclei in the red boxes of (A) and (B) is 36.5 %, and 36.8 % between the orange boxes in (B) and (C). The relative dynamics intensity difference is only 4.9 % between (A) and (B) and 3.3 % between (B) and (C).

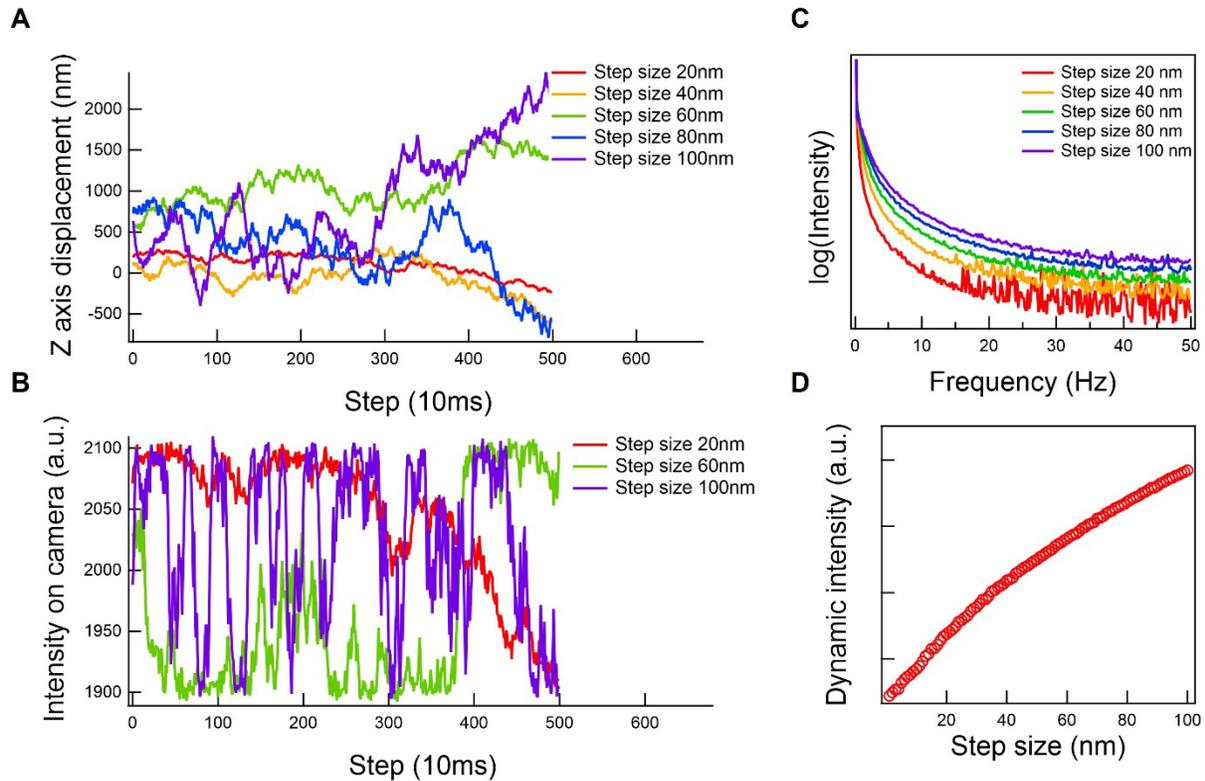

**Figure S3. Simulation of dynamic intensity based on random walk model.** (**A**) Z-axis displacement of particles with different step sizes of random walk. In each random walk, the angle between the direction of the random walk and the z-axis is random. Since the sampling rate of the camera in the experiment is 100 frames per second and the sampling time is 5 seconds, the time interval of each random walk simulated here is 10 milliseconds, and a total of 500 random steps are walked. (**B**) Variation of interference signals caused by random walk of scatterers. Poisson noise has been added to the light intensity. (**C**) Frequency spectra obtained by Fourier transform the light intensity traces in (B), each spectral curve is obtained by an average of 1000 times. (**D**) Relationship between step size of random walk and dynamic intensity (spectral integral in (C)). All scatterers are set to have the same reflectivity. Since the step frequency is constant, the random walk step size reflects the velocity of scatterers.

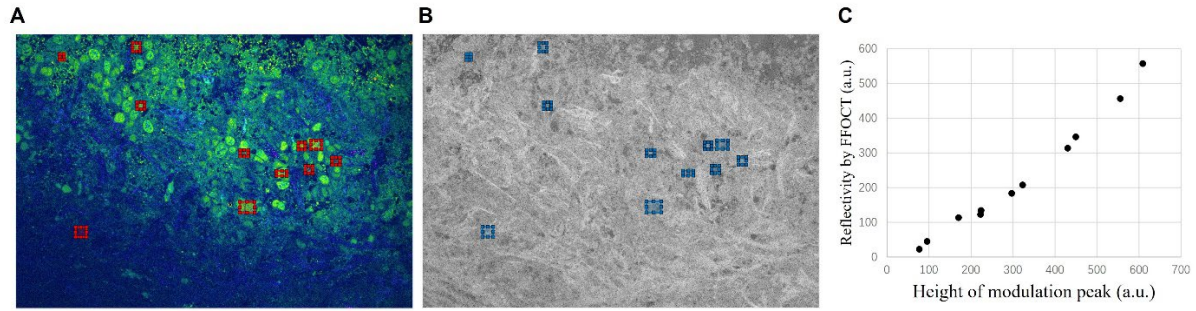

**Figure S4. Comparison of modulation peak height and FFOCT image intensity in the same region.** (**A**) APMD-FFOCT image, where red boxes are randomly selected regions. (**B**) FFOCT image of the same tissue in (A), where blue boxes mark the same area as red boxes in (A). The intensity in the FFOCT image represents their reflectivity. (**C**) The relationship between the height of the modulation peak extracted from red boxes in (A) and the reflectivity intensity extracted from blue boxes in (B).

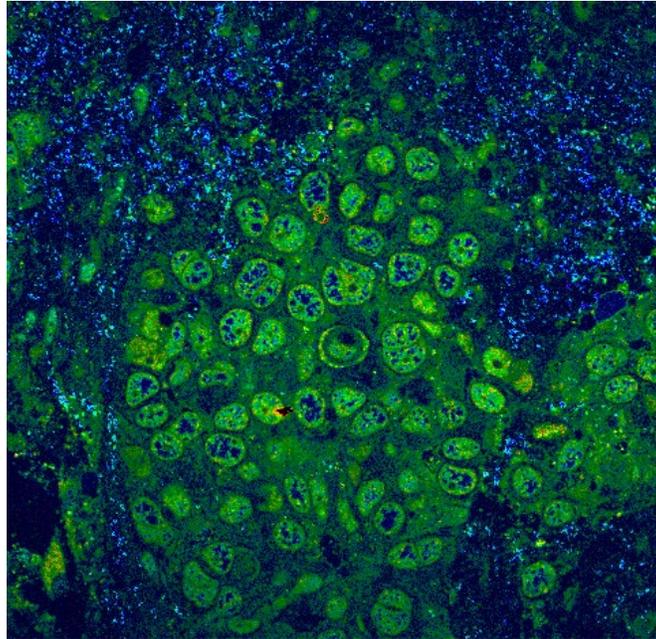

**Figure S5. The combination image of Fig. 2a and Fig. 2b.** Both collagen fibers and tumor cells are clearly presented in this figure. Moreover, the brightness and hue of collagen fibers can be adjusted independently and flexibly.

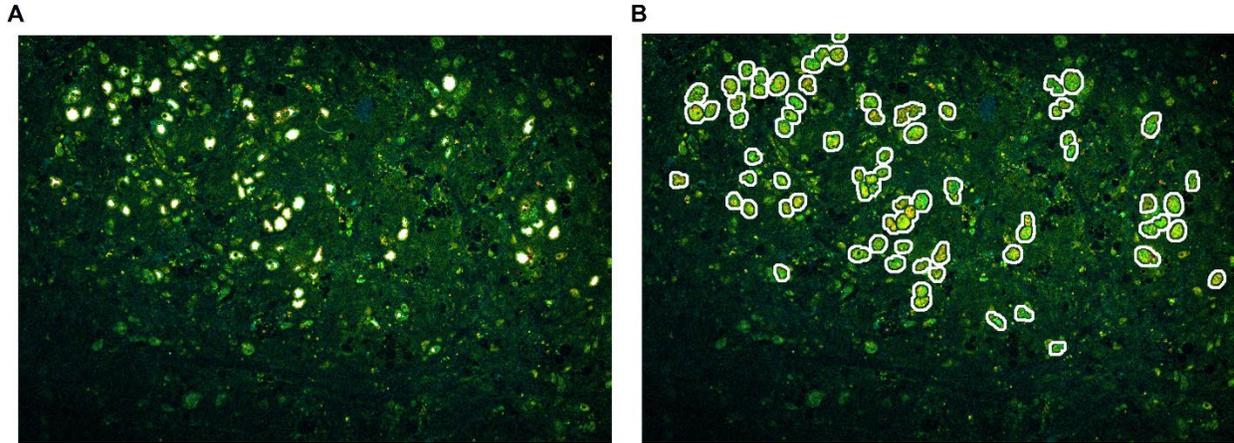

**Figure S6. Automatic labeling of nuclear and cytoplasmic regions.** (**A**) The white areas mark the nuclei, wherein the nucleoli within the cell nuclei have been filtered out. The connected or overlapping nuclei are separated. (**B**) The white areas mark the cytoplasm, surrounding the nuclei. The automatic labeling operations are repeated in each subgraph of stitched APMD-FFOCT images.

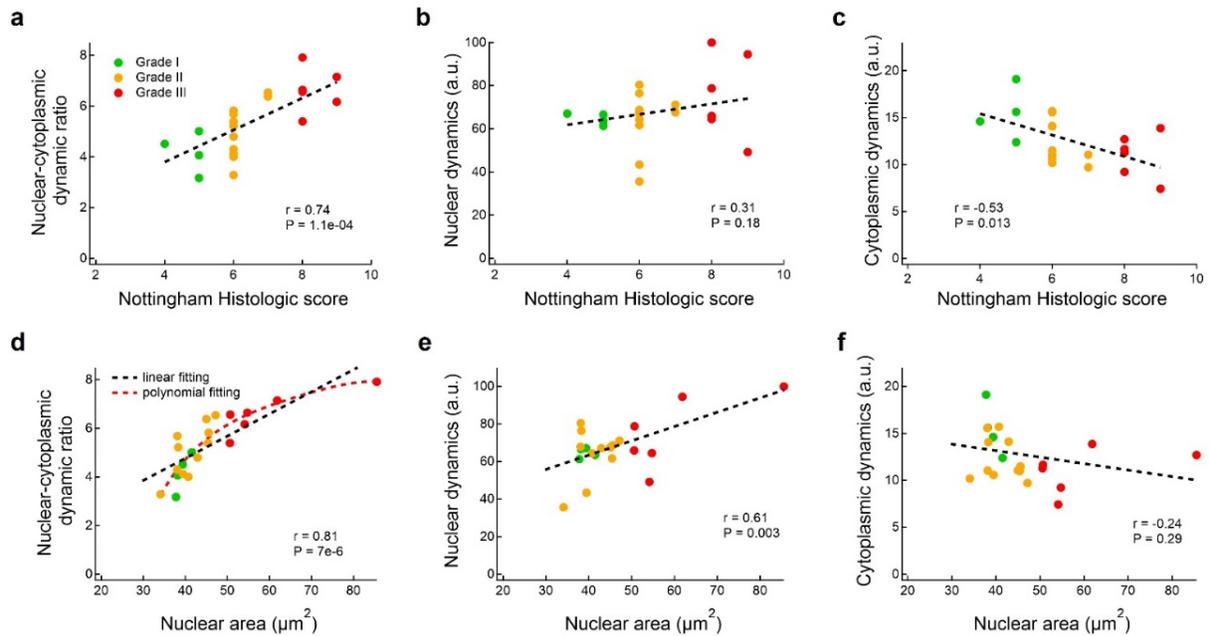

**Figure S7. The dynamic intensity's dependency on tumor grading and nuclear area, as well as the fitting results.** (**A-C**) The dependence of nuclear-cytoplasmic dynamic intensity ratio (A), nuclear dynamic intensity (B), and cytoplasmic dynamic intensity (C) on the Nottingham Histologic Score. (**D-F**) The dependence of nuclear-cytoplasmic dynamic intensity ratio (D), nuclear dynamic intensity (E), and cytoplasmic dynamic intensity (F) on the nuclear area. The red dotted line in (D) is the polynomial fitting, which indicates a more nonlinear relationship between NCDR and the nuclear area. Each point represents averaged data from an individual sample (Table S1), categorized by tumor grade as indicated by the color coding: Grade I (green), Grade II (orange), and Grade III (red). The black dotted lines represent the fitted linear regression models, where "r" represents the Pearson correlation coefficient, and "P" signifies the significance level of the correlation.

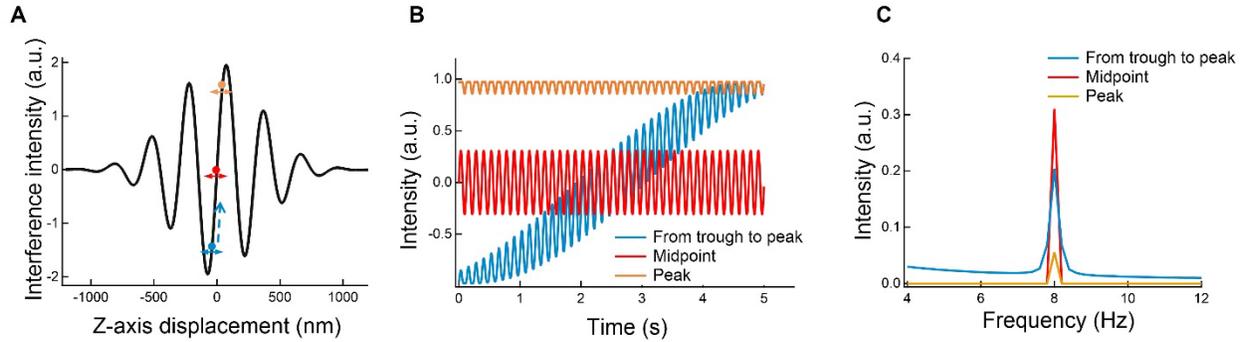

**Figure S8. Sensitivity of interferometry response to phase variations in scatterers.** (**A**) Change in interference intensity as a function of scatterer displacement along the z-axis. The three balls represent scatterers in different phases, in which the red and orange balls are at the midpoint and the peak respectively, the blue balls move from the trough to the peak, and the double arrows below the balls indicate that the balls vibrate near their phase under the same active phase modulation. (**B**) The intensity fluctuations resulting from active phase modulation across different interference phases, correspond to the three balls in (A). Among them, the amplitude of intensity fluctuations caused by the active modulation of the red ball in the middle point is the largest, and the amplitude of intensity fluctuations caused by the active modulation of the orange ball near the peak is the smallest, while the amplitude of intensity fluctuations varies with the phase when the blue ball moves from the trough to the peak. (**C**) The active modulation peaks on spectra correspond to three balls, in which the active modulation peak of the orange ball is the smallest, the active modulation peak of the red ball is the largest, and the modulation peak of the blue ball is between the two, and its height reflects the average effect of active modulation during its movement. The results show that the response sensitivity of interferometry is related to the interference phase of the scatterers, and the active phase modulation can reflect the impact of phase factors for both stationary or moving scatterers.

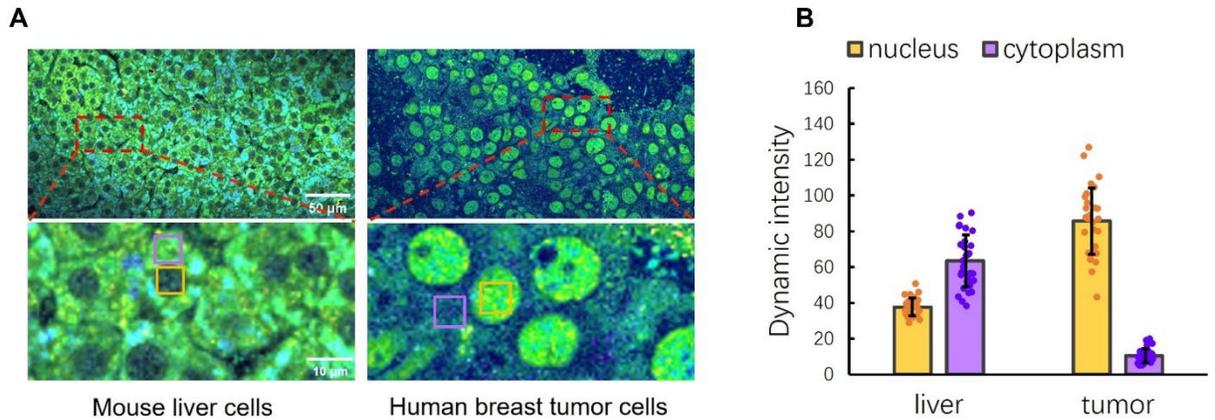

**Figure S9. Comparison of dynamic intensity of nuclei and cytoplasm of liver and breast tumor cells.** (**A**) The left column features mouse liver cells, where the nuclei appear darker and the cytoplasm is brighter. On the right are human breast tumor cells (IDC, grade III), displaying the opposite pattern: the nuclei are lighter while the cytoplasm is darker, contrasting with the liver cells. (**B**) The dynamic intensities of the cytoplasm and nuclei in (A) were analyzed. In liver cells, the nuclear intensity exceeds that of the cytoplasm, which contrasts with tumor cells. Moreover, the dynamic intensity of the tumor cell nuclei is higher than that of liver cells, corresponding to their increased proliferation rate. Meanwhile, the dynamic intensity of the liver cell cytoplasm is greater than that of tumor cells, corresponding to the complex and important metabolic activities occurring within the liver cell cytoplasm. Error bars: ±1 SD. N = 29 in liver and N = 31 in tumor.

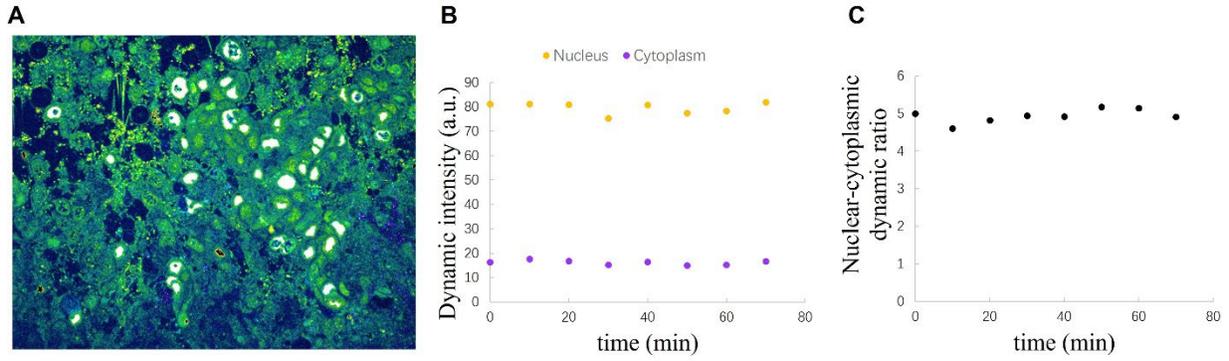

**Figure S10. Temporal Analysis of dynamic intensity.** (**A**) APMD-FFOCT image highlighting cell nuclei in white for dynamic analysis. (**B**) Dynamic intensity of cell nuclei (yellow) and cytoplasm (purple) over 70 minutes. (**C**) Nuclear-cytoplasmic dynamic ratio over 70 minutes.

| Sample | Cell number | Mean NCDR | Mean nuclear area/μm | Nuclear dynamics | Cytoplasmic dynamics | Diagnosis |
|--------|-------------|-----------|----------------------|------------------|----------------------|-----------|
| 1 | 148 | 3.17 | 37.77 | 61.29 | 19.10 | 2+2+1 Grade Ⅰ,IDC |
| 2 | 6295 | 5.01 | 41.48 | 63.51 | 12.38 | 2+2+1 Grade Ⅰ,IDC |
| 3 | 2542 | 4.52 | 39.39 | 67.04 | 14.61 | 1+2+1 Grade Ⅰ,IDC |
| 4 | 1881 | 4.07 | 38.15 | 66.54 | 15.62 | 2+2+1 Grade Ⅰ,IDC |
| 5 | 1140 | 4.00 | 40.68 | 64.34 | 15.70 | 3+2+1 Grade Ⅱ,IDC |
| 6 | 658 | 4.13 | 39.46 | 43.38 | 10.58 | 3+2+1 Grade Ⅱ,IDC |
| 7 | 2511 | 6.38 | 44.99 | 67.49 | 11.05 | 3+2+2 Grade Ⅱ,IDC |
| 8 | 4040 | 3.28 | 34.08 | 35.61 | 10.19 | 3+2+1 Grade Ⅱ,IDC |
| 9 | 809 | 4.30 | 38.07 | 67.84 | 15.58 | 3+2+1 Grade Ⅱ,IDC |
| 10 | 290 | 5.68 | 38.11 | 80.41 | 11.05 | 2+2+2 Grade Ⅱ,IDC |
| 11 | 606 | 6.54 | 47.15 | 71.13 | 9.72 | 3+3+1 Grade Ⅱ,IDC |
| 12 | 2281 | 5.82 | 45.56 | 68.70 | 11.50 | 3+2+1 Grade Ⅱ,IDC |
| 13 | 919 | 5.39 | 45.41 | 61.71 | 10.98 | 3+2+1 Grade Ⅱ,IDC |
| 14 | 1315 | 4.80 | 42.91 | 67.08 | 14.11 | 3+2+1 Grade Ⅱ,IDC |
| 15 | 1717 | 5.22 | 38.26 | 76.40 | 14.06 | 3+2+1 Grade Ⅱ,IDC |
| 16 | 1859 | 7.92 | 85.53 | 99.89 | 12.70 | 3+3+2 Grade Ⅲ,IDC |
| 17 | 640 | 5.40 | 50.59 | 65.90 | 11.27 | 3+3+2 Grade Ⅲ,IDC |
| 18 | 1364 | 6.57 | 50.67 | 78.73 | 11.66 | 3+3+2 Grade Ⅲ,IDC |
| 19 | 818 | 7.15 | 61.85 | 94.46 | 13.87 | 3+3+3 Grade Ⅲ,IDC |
| 20 | 335 | 6.17 | 54.12 | 49.15 | 7.42 | 3+3+3 Grade Ⅲ,IDC |
| 21 | 178 | 6.64 | 54.74 | 64.44 | 9.22 | 3+3+2 Grade Ⅲ,IDC |

**Supplementary Table S1.** Table summarizing pathology results of 21 samples in Fig. 4l.